\documentstyle[prl,aps,twocolumn,epsfig]{revtex}
 \newcommand{\be}{\begin{equation}}
 \newcommand{\ee}{\end{equation}}
 \newcommand{\bdis}{\begin{displaymath}}
 \newcommand{\edis}{\end{displaymath}}

\begin{document}
 \bibliographystyle{prsty}  
 
 \title{ Analytic calculation of anomalous scaling 
 in random shell models for a passive~scalar}
 \author{ R. Benzi$^{1}$, L. Biferale$^{2,3}$,  and A. Wirth$^{4}$ }
 
 \date{\today}
 \address{$^1$ AIPA, Via Po 14, 00100 Roma, Italy\\
 $^2$  Dept. of Physics, University of Tor Vergata,
 Via della Ricerca Scientifica 1, I-00133 Roma, Italy\\
$^{3}$ Istituto Nazionale di Fisica della
 Materia, unit\`a di Roma Tor Vergata\\
 $^{4}$ Observatoire de la C\^ote d'Azur, B. P. 229, 06304 Nice, France}
 \maketitle 
 \begin{abstract}
 An exact non-perturbative calculation of the fourth-order anomalous
  correction  to the scaling behavior
 of a  random shell-model for passive scalars is presented. 
 Importance of ultraviolet (UV)  and infrared 
 (IR) boundary conditions
 on the inertial scaling properties are determined.
 We find that anomalous behavior is 
 given by the null-space of the inertial operator and we prove strong 
 UV and IR
 independence of the anomalous exponent. A limiting case where diffusive 
 behavior can influence inertial properties is also presented.
 \end{abstract}

 Recently, there has been growing evidence, both numerically and
 experimentally, that fully developed turbulence is characterized by 
anomalous
 scaling of the velocity structure functions 
$F_p \equiv \langle  |v(x+r)-v(x)|^p \rangle$.
 In particular, it has been shown that,
 within the inertial range (i.e. $\eta \ll r \ll L $,
 $\eta$ being the Kolmogorov dissipation scale and $L$ the outer length of
 external forcing) $F_p(r) \sim r^{z_p}$, where $z_p$ is a non linear function
 of $p$ which do not follow dimensional
 counting, i.e. $z_p \neq \frac{p}{q} z_q$.
 One of the most challenging scientific issue is to develop a theory which
 allows a systematic computation of $z_p$ by using the equation of motions.
 
 Recently \cite{K94,GK,CFL,P,E} 
 this issue has been addressed
 by studying a highly non trivial "toy model" introduced by Kraichnan,
 namely the advection
 of a passive scalar by a random, Gaussian velocity field,
 white-in-time and whose two-point
velocity correlation function is given by:$\langle v_i(x,t)
v_j(x',t')\rangle=\delta(t-t')D_{ij}(|x-x'|)$, with $D_{ij}(x) =
D_{ij}(0)-\hat{D}_{ij}(x)$. Here,
  $\hat{D}$ is the $d$-dimensional velocity-field
structure function\,:
$
\hat{D}_{ij}(x) = D_0 |x|^{\xi} \left[ (d-1+\xi)\delta_{ij}
 - \xi x_ix_j |x|^{-2} \right],
$
where the scaling exponent of the second order velocity
structure function, $\xi$ ($0 \ll \xi < 2$), is a free parameter.
Higher order velocity-field correlation functions are fixed by
the Gaussian assumption.
 
 Although such a choice is far from being
 realistic,
 many interesting  analytical and phenomenological results 
 have been obtained for this 
 toy-model. Due to the delta correlation in time, moment equations
to all orders are closed. 
 In \cite{K94,K68}, for the first time Kraichnan  gave  the
 closed expression for $\tilde{S}_2(r)$, where 
 \begin{equation}
 \tilde{S}_p(r)  \equiv  \langle|\theta(x)-\theta(x+r)|^p\rangle 
 \sim  r^{\zeta_p}.
 \label{eq:sf}
 \end{equation}
 $\theta(x)$ being the scalar field transported by the turbulent velocity 
 field. In the inertial range 
 $\tilde{S}_p(r) \sim  r^{\zeta_p} $ and the
 set of scaling exponents $\zeta_p$ fully characterizes
 intermittency.
 In \cite{K94} a
 theory for all structure functions is proposed and an explicit
 formula for $\zeta_p$ is derived.   
 The main physical outcome is that
 all structure functions of order greater than two have intermittent
 corrections and that intermittency should be connected to some non
 trivial matching between advective and diffusive properties of the
 model. 
 
 In \cite{GK,CFL} it has been shown that intermittency of the
 scalar structure functions is connected to the properties of the
 null-space of the linear operator appearing in the equation of 
 multipoint passive scalar moments. Moreover, in \cite{GK,CFL}
  a perturbative
 expression of intermittency correction, as a function of the
 parameter $\xi$ and of the system dimensionality has been derived.
 Both in \cite{K94} and in \cite{GK,CFL},
 matching conditions at infrared (IR) and
 ultraviolet (UV) scales should be taken into account. 
The same problem for the case of passive vectors  was 
 addressed in \cite{massimo}. 
 Finally, in all cases, universality in
 the scaling exponents is supposed to be preserved. 
 
 In \cite{wb} two of us have  worked  out an even simpler
 toy-model which displays connections to the physics of
 a passive scalar advected by a random velocity field,
 being at the same time   more tractable both
 analytically and numerically.
 In \cite{wb} the
 intermittent properties of a shell-model for a passive scalar
 advected by a delta-correlated  random velocity field have
 been investigated. The main results presented in \cite{wb} were 
 that:
  (i) the
 second order structure function has no anomalous scaling,
 i.e. $ \zeta_2 = 2 - \xi$; 
 (ii) all structure
 functions of order larger than two have anomalous corrections;
  (iii) anomalous behavior
 tends to vanish when approaching the laminar regime, $\xi=2$.
 For all the three points the model agrees with the Kraichnan model.
 
 \noindent
 The importance of UV and IR boundaries for  the anomalous
 scaling was left unanswered in \cite{wb}. In particular,
 numerical simulations were unable  to 
 distinguish among contribution coming from the inertial null-space and
 possible singular behavior introduced by the boundary conditions.
 
 \noindent
 In this letter we show how to compute exactly and 
 non-perturbatively the inertial scaling behavior of the fourth-order
 structure functions. The main result is that the scaling
 properties are completely dominated by the null-space of the
 inertial linear operator and strongly universal.
 The signature of UV and IR cut-offs is due to the presence
 of sub-dominant terms which weakly perturb the 
 pure-scaling  behavior of the inertial operator.
 Anomalous scaling is calculated for any $\xi > 0$.
 We also present some results which support the strong singular nature
 of the limit $\xi \rightarrow 0$,
 In this limit, due to the non-local
 nature of interactions,
 it is not possible to neglect UV effects   
 if the diffusive scale, $k_d$, is taken fixed.
 
 We first introduce a simplified version of
 the shell model discussed in 
 Ref.~\cite{wb}.
 The model is defined in terms of a shell-discretization
 of the Fourier space in a set of wavenumber defined on a geometric
 progression $k_n =  \lambda^n$, with $\lambda > 1$. 
Passive increments at scale $r_n=k_n^{-1}$
 are described by a complex variable $\theta_n(t) $. The
 time evolution is obtained according to the following criteria\,: (i)
 the linear term is a purely diffusive term given by $-\kappa k_n^2 
 \theta_n$;
 (ii) the advection term is a combination of the form
  $k_n \theta_{n'} u_{n''}$; (iii) interacting shells are restricted to
 nearest neighbors of $n$; (iv) in the absence of forcing
 and damping the model preserves the volume in the phase-space and 
 the passive-energy $ E = \sum_n |\theta_n|^2$. Properties (i), (ii) and 
 and (iv) are valid also for the original equation of a passive
 scalar advected by a Navier-Stokes velocity field
 in the Fourier space, while property (iii) is an assumption
 of locality of interactions among modes. This assumption is rather well
 founded as long as  $0 \ll \xi < 2$. 
  The model 
 is defined by the following equations ($m=1,2,\ldots$)
 \begin{eqnarray}
 [\frac{d}{dt} + \kappa k_m^2] \theta_m (t) =& \nonumber \\ 
     i [c_{m} 
 \theta_{m+1}^* (t)  u_{m}^*(t)
 + &b_m \theta_{m-1}^* (t) u_{m-1}^* (t)) +\delta_{1m} f(t)
 \label{shellmodel}
 \end{eqnarray}
 where the star denotes complex conjugation and $ 
  b_{m} = -k_{m},
  c_{m} =k_{m+1}$ for imposing energy conservation in the zero diffusivity
limit.
 Boundary conditions are defined as: $u_0=\theta_0=0$. 
 The forcing term $\delta_{1m} f(t)$is gaussian and delta correlated:
$\angle f(t)f(t') \rangle =F_1 \delta(t-t')$  
 acts only on the first shell. 
 In numerical implementations, the model is truncated to a
  finite number of shells $N$  with
  the additional boundary conditions 
 $\theta_{N+1}=0$.
 
 \noindent
 Following Kraichnan \cite{K68} we assumed  that the
 velocity variables $u_m (t)$ and the forcing term $f(t)$ are
 independent complex Gaussian and white-in-time, with scaling law:
 $\langle u_m (t) u_n^* (t') \rangle = \delta(t-t') \delta_{nm} d_m, \, \, 
 d_m= k_m^{-\xi}$
 Due to the delta-correlation in time,
 we can close the  equations of motion for all  structure functions.
 Numerical simulations show that the model has
 the same qualitative intermittency
 of the model studied in \cite{wb}.
 
 In this letter we concentrate on the non-perturbative analytic 
 calculation of the  fourth-order structure 
 function
 $P_{mm}=  \langle (\theta_m \theta_m^*)^2 \rangle \propto k_m^{-\zeta_4}$
 (the lowest order with non-trivial anomalous scaling).
 The closed
 equation satisfied by $P_{mq}=\langle
(\theta_m \theta_m^*)(\theta_q \theta_q^*)\rangle$,
 is,
 \begin{eqnarray}
 \dot{P}_{mq} =  (\delta_{1,m} &E_{m}&+\delta_{1,q} E_{q}) F_1  
 -  \kappa (k_m^2 +k_q^2) P_{mq}   \nonumber \\
 +[-P_{mq}  c_m^2d_m&(
 (1&+\delta_{q,m+1}) + \lambda^{\xi-2} 
  (1+\delta_{q,m-1}) ) \nonumber  \\
 + P_{m+1,q} c_m^2& d_m (1&+\delta_{q,m})+ P_{m-1,q} b_m^2 d_{m-1}
 (1+\delta_{q,m})\, \, \, \nonumber \\
 +( q \leftrightarrow m) ] && \; \; \; \; 
 \label{4}
 \end{eqnarray}
where $E_n = \langle\theta_n\theta_n^* \rangle$. 
 We can symbolically represent eq. (\ref{4}) as:
 \begin{equation}
 \dot{P}_{mq} = {\cal I}_{mq,lp}  P_{lp} +\kappa { \cal D}_{mq,lp} P_{lp}
 + {\cal F}_{mq}
 \label{eqgen}
 \end{equation} 
 where  ${\cal I }$ and  ${\cal D}$ are the  inertial and the
 diffusive 4-order tensor and  ${\cal F}$ is the forcing term. 
 
 %%%%%%%%%%%%%%%%%%%%%%%%%%%%%%%%%%%%%%%%%%%%%%
 
 \noindent
 Our main result is derived by using the following ansatz: the
 {\it symmetric}
 matrix $P_{mq}$, which fully determines the
 scaling properties for any fourth-order quantity in the model, can be 
 described as:
 \begin{equation}
 P_{n,n+l} = C_l P_{n,n} \,\, (l \ge 0), \; \; \;
 P_{n,n-l} = D_l P_{n,n}\,\,  (l \ge 0)
 \label{scalingmatrix}
 \end{equation}
 The independency of $C_l $ and $D_l$ from $n$ is equivalent to demand 
 absence 
 of  strong boundary effects, i.e. the matrix is formally
 infinite-dimensional. Clearly this must be verified a posteriori
 showing that the solution we are going to present is UV and IR stable.
 
 \noindent
 Using (\ref{scalingmatrix}) we obtain:
 $$
 \frac{C_{l+1}}{C_l C_1} = \frac{D_{l+1}}{D_l D_1}
 $$
 which is equivalent to write: $P_{n+l,n+l} = k_l^{-\zeta_4} P_{n,n}$
 where  $
 C_l/D_l = k_l^{-\zeta_4}$ and $  \zeta_4 = 2(2-\xi) -\rho_4. $ 
 As usual we indicate by $\rho_4$ the anomalous correction to the scaling
exponent.
 
 Let us notice that (\ref{scalingmatrix}) does not force 
the solution
 to have global scaling invariance: only the diagonal part is requested to have
 pure scaling.
 
 Let us proceed by analyzing (\ref{4}) restricted to
 the inertial operator and for the diagonal ($m=q$) and  sub-diagonal 
 terms ($q=m-1$):
 \begin{eqnarray}
 \dot{P}_{m,m}&  =& 2 P_{m,m} c^2_m d_m (-1 -x +2( C_1 +D_1 x) )  \\
 \dot{P}_{m,m-1}&  =& 2 P_{m,m-1} c^2_m d_m (-1 -4x  -x^2 \nonumber \\
 +\frac{x}{D_1} & +&\frac{x+C_2+\frac{x^{2}C_2}{R}}{C_1}) 
 \label{l0l1}
 \end{eqnarray}
 where we have posed  $x=\lambda^{\xi-2}$
 and $R=C_1/D_1$.
 By plugging the scaling (\ref{scalingmatrix}) in (\ref{l0l1})  one obtains
 two equation in three unknowns which can be taken to be $C_1$ and the ratios
 $C_2/C_1$ and $R$.
Numerical investigation suggest the following
"scaling ansatz": 
\begin{equation}
P_{n,n+l} = C_l  P_{n,n},\, \mbox{ with} \,\,  C_l = C_1 k_{l-1}^{\xi-2}
\end{equation}
and
\begin{equation}
P_{n,n-l} = D_l  P_{n,n},\,\, \mbox{  with} \,\, 
D_l=D_1 k_{l-1}^{-(\xi-2)-\rho_4}
\end{equation}
where anomalous correction is felt only in the IR part of the matrix.
By plugging this scaling in (\ref{l0l1})  we end up with two eqs. in
two unknowns and we can calculate $\rho_4$.
 Let us anticipate that this 
 (wrong) assumption gives  results in very good agreement with the numerical 
 simulations, indicating that the true solution is not very far from
 having pure scaling behavior.
 In order to solve the full problem, without imposing any
 "pure scaling" behavior, we analyze the other
 entry of the matrix $P_{n,q}$ with $q \neq n $ and $ q \neq n-1$.
 Let us put  $\gamma_l  = D_{l+1}/D_{l}, \delta_l = C_{l+1}/C_l$.
 
 It is then possible to show  that
 for $l > 1$ by plugging the scaling (\ref{scalingmatrix}) in the inertial part 
 of (\ref{4}) and studying the 
 equation for $\dot{P}_{n,n\pm l}$ we obtain two recursion equations: 
 \begin{eqnarray}
 -(1+x)(1+x^l) +\gamma_l(R+x^{l+1}) +\frac{1}{\gamma_{l-1}} (x^l+
 \frac{x}{R} )& =&0
 \label{recursion} \\
 (1+x)(1+x^l) -\lambda^{\zeta_4} \delta_l(R+x^{l+1}) -\frac{\lambda^{-\zeta_4}}
 {\delta_{l-1}} (x^l+\frac{x}{R}  )& =&0
\label{recursion2}
 \end{eqnarray}
These two relations can indeed be seen as two maps connecting 
successive values of $\gamma_l$ and $\delta_l$ respectively. By iterating
forward (backward) the map (\ref{recursion}) we move from the diagonal
(IR boundary) to the IR boundary (diagonal) along a row of the matrix $P_{l,n}$.
 By iterating 
forward (backward)
the map  (\ref{recursion2})  we move from the diagonal
(UV boundary) to the UV boundary (diagonal) along a row of the matrix $P_{l,n}$.

 Let us first note that the two maps are not independent, i.e. they
satisfy our scaling ansatz
 $\delta_l = R \gamma_l$ and therefore we are going to consider only one of the
 two  in  what follows. 
 In order to test stability under weak perturbation
of boundary conditions in the map ({\ref{recursion}) 
we are  interested in the behavior 
 by backward iterations, i.e. iterating from $l = \infty $ to $l=0$.
 In the limit $l 
 \rightarrow \infty$ and $\xi \neq 2$  the map (\ref{recursion})
 has only two fixed point corresponding to $ \gamma_1^* = x/R$ and 
 $\gamma_2^* = 1/R$. It turns out that $\gamma_1^*$ is stable for back 
 iterations, i.e. iterating from the 
 IR boundary ($l \gg 1$) to the diagonal ($l=0$). The global solution can now
 be obtained by  
 a self-consistent method. First, let us take as initial value for $R$ the 
  value that one would have guessed from  imposing "pure scaling" as discussed 
 previously, then we can iterate (\ref{recursion}) from the boundary toward the 
 diagonal and finding the value for $C_2/C_1$. This value can be used 
 to close (\ref{l0l1}) exactly. Next, with the improved value for $R$,
 one can restart the full procedure getting a new improved value of $R$ and
 so on  up to the moment when the new value of $R$ reach it fixed point.
 
\noindent
 In figure 1 we show the computation of $\rho_4$ obtained by numerical 
 integration of equation (\ref{shellmodel}) as a function of $\xi$. 
In the same figure
 we plot the $\rho_4$ as a function of $\xi$ obtained by the analytical
 solution previously discussed.
  As one can see, the agreement  is perfect. Let us notice that it is
impossible to go by numerical simulations to values of $\xi$ very near
zero because of strong diffusive effects which completely destroy scaling
behavior.

\noindent
 Let us notice that it is the strong 
 stability under 
 UV and IR perturbations  that allows us to iterate consistently the 
 procedure. 
 We have therefore proved that anomalous scaling comes only
 from the inertial operator and that it shows a very 
 strong degree
 of universality as a function of the forcing and dissipative
 mechanisms, at least as far as the situation with $\xi =const. > 0$ 
 and 
 molecular diffusivity $\kappa \rightarrow 0$ is considered. 
 
 \noindent 
 Some new and interesting phenomenum happens when we are in the other 
 possible asymptotic
 limit ($ \xi \rightarrow 0$) and fixed molecular diffusivity. A simple 
 dimensional argument
 tell us that the following relation holds:
 \begin{equation}
 \kappa = const. k_d^{-\xi},
\label{diss}
\end{equation}
 where with $k_d$ we mean a preassigned diffusive scale such that all the 
 inertial dynamics is at $k \ll k_d$.  One can show  that in such a 
 situation the diffusive operator ${\cal D}$ 
 in (\ref{eqgen}) gives  a contribution of the form 
 \begin{equation}
 \dot{P}_{n,n} =- const.  (k_n/k_d)^{\xi} k_n^{2-\xi} P_{n,n} 
 \end{equation}
 Therefore, the diffusive perturbation is absolutely 
 negligible in
 the case when $\xi \neq 0 $ is fixed and $k_d \rightarrow \infty$ while 
 it becomes
 a singular perturbation when $k_d$ is kept fixed and $\xi \rightarrow 0$. 
 A detailed
 analysis of relations (\ref{l0l1})
 shows that in the singular limit $\xi \rightarrow 0$ there 
 appear an infinitesimal interval
 of values of $\xi \sim 0$ where $\rho_4 \rightarrow 0$, i.e. the 
 anomalous correction
  tends to vanish with a particular shape which depends on the constants
appearing in relation (\ref{diss}).
 
 \noindent 
 This concludes the presentation of our results.

\noindent
 Let us remark that the 
 perfect agreement 
 of our inertial null-space solution with the numerical simulation 
 performed with finite diffusivity and in  presence of forcing
 is the clear demonstration that the scaling behavior is completely 
 dominated by the inertial operator. For any finite system,  IR and 
 UV effect weakly perturb the pure
 scaling solution.
  Our result  shows that the inertial operator is perfectly suitable for 
 picking all anomalous
 aspects but in the case where strong non local interactions 
(dynamically produced)  completely 
 destroy inertial properties introducing diffusive effects at all scales 
 ($\xi \rightarrow 0$ and $k_d$ fixed). 

Discussions with B. Dubrulle, G. Falkovich, U. Frisch, I. Procaccia
 and M. Vergassola are kindly
acknowledged. 

\bigskip
{\bf Figure caption}

\noindent
Fig1: analytical ansatz (continous line) and numerical results (squares)
for $\rho_4$ 
are plotted for various values of $\xi$.

 
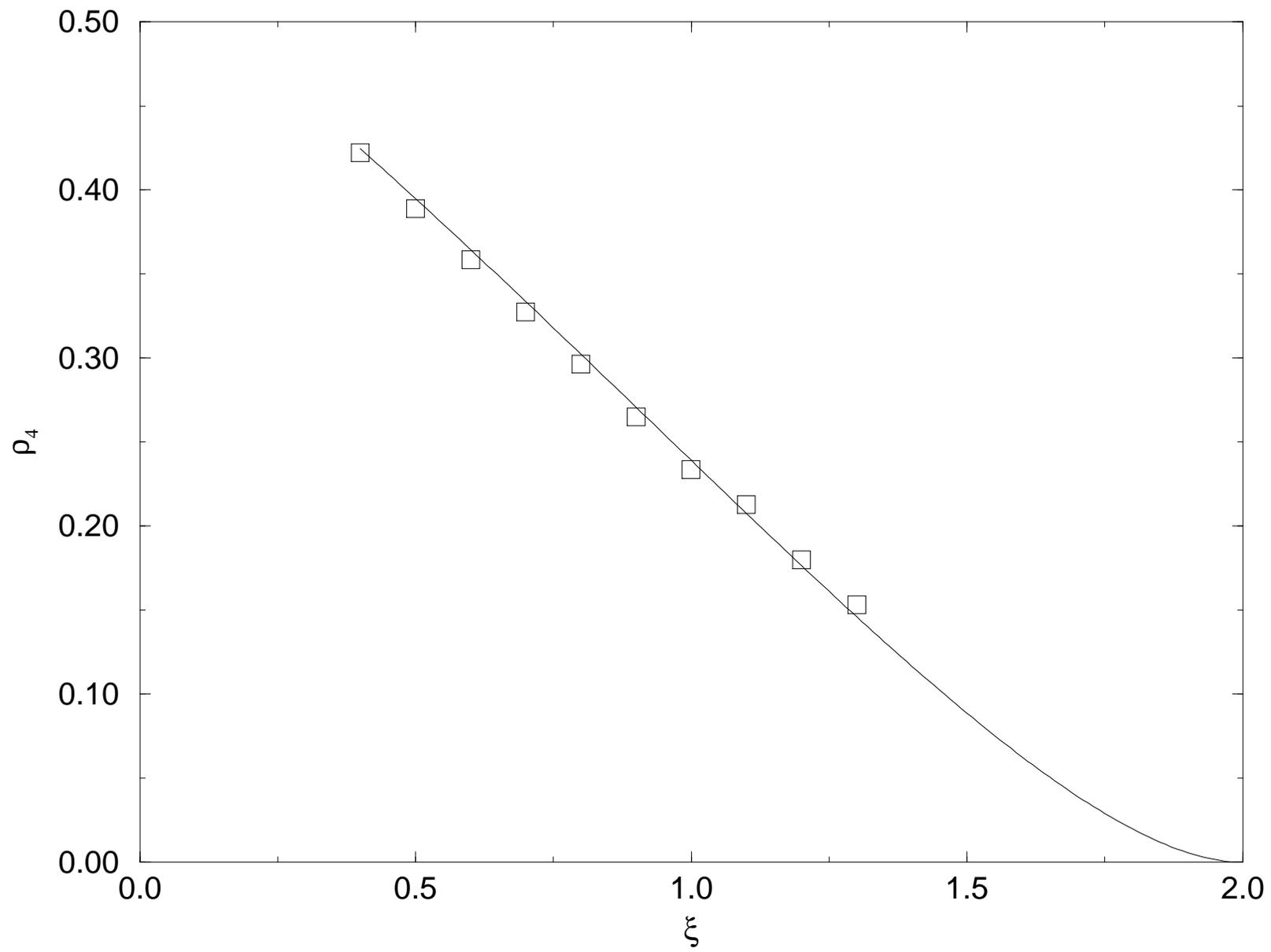
\end{document}